\documentclass[10pt,conference]{IEEEtran}
\IEEEoverridecommandlockouts

\usepackage{cite}
\usepackage{booktabs}
\usepackage{multirow}
\usepackage{amsmath,amssymb,amsfonts}
\usepackage{graphicx}
\usepackage{textcomp}
\usepackage{xspace}
\usepackage[table]{xcolor}
\usepackage{float}
\usepackage{caption}
\usepackage{listings}
\usepackage{stfloats}
\usepackage{enumitem}
\usepackage[utf8]{inputenc}
\usepackage[most]{tcolorbox}
\usepackage{comment}
\usepackage{url}

\definecolor{findinggreen}{RGB}{147, 179, 148}
\definecolor{bordergrey}{RGB}{200, 200, 200}

\newtcolorbox{findingbox}[1]{
    enhanced,
    before skip=2mm,
    after skip=2mm,
    colback=white,
    colframe=bordergrey,
    boxrule=0.5pt,
    sharp corners,
    attach boxed title to top left={
        xshift=5mm, 
        yshift=-2mm, 
        yshifttext=-1mm
    },
    boxed title style={
        colback=findinggreen,
        colframe=findinggreen,
        sharp corners,
        left=2pt, right=2pt, top=1pt, bottom=1pt,
        boxrule=0pt
    },
    fonttitle=\bfseries\sffamily\small,
    coltitle=white,
    title={Finding #1},
    left=2mm, right=2mm, top=1mm, bottom=1mm
}

\tcbset{
  promptbox/.style={
    enhanced,
    colback=gray!5,
    colframe=gray!50,
    boxrule=0.5pt,
    arc=2mm,
    outer arc=2mm,
    fonttitle=\bfseries,
    coltitle=black,
    title={#1},
    sharp corners=south,
    breakable,
    listing only,
    listing options={basicstyle=\ttfamily\footnotesize, breaklines=true}
  }
}

\definecolor{keycolor}{rgb}{0.0, 0.0, 0.8}
\definecolor{valuecolor}{rgb}{0.8, 0.1, 0.1}

\def\BibTeX{{\rm B\kern-.05em{\sc i\kern-.025em b}\kern-.08em
    T\kern-.1667em\lower.7ex\hbox{E}\kern-.125emX}}

\newcommand{\approach}{{SkillForge}\xspace}

\newcommand{\up}[1]{{\small\textcolor{green!60!black}{$\uparrow$#1}}}
\newcommand{\upbf}[1]{{\small\textbf{\textcolor{green!60!black}{$\uparrow$#1}}}}
\newcommand{\down}[1]{{\small\textcolor{red!70!black}{$\downarrow$#1}}}

\newcommand{\sigone}{{\small\textsuperscript{\dag}}}

\begin{document}
\title{\approach: Self-Distilling Agents for Project-Specific Issue Resolution}
\author{
\IEEEauthorblockN{Silin Chen\IEEEauthorrefmark{1}, Han Li\IEEEauthorrefmark{1}, Xiaodong Gu\IEEEauthorrefmark{2}, Yuling Shi, Haibing Guan}
\IEEEauthorblockA{\textit{Shanghai Jiao Tong University}\\
\texttt{cslsolow@gmail.com, lihan0421@sjtu.edu.cn, xiaodong.gu@sjtu.edu.cn}\\
\texttt{yuling.shi@sjtu.edu.cn, hbguan@sjtu.edu.cn}}
\thanks{\IEEEauthorrefmark{1}Silin Chen and Han Li contributed equally to this work.}
\thanks{\IEEEauthorrefmark{2}Corresponding author: Xiaodong Gu.}
}

\maketitle

\begin{abstract}
Large language model (LLM) based agents have demonstrated remarkable proficiency in automated software issue resolution, yet they often struggle to resolve issues in a specific repository because they lack project-specific knowledge. Existing self-evolving approaches acquire such knowledge from repository history or online repair trajectories, but they either depend on available historical issue-resolution signals or incur substantial per-issue test-time exploration cost.
In this paper, we propose \approach, a self-distillation framework that proactively acquires project-specific knowledge from the repository itself. Instead of waiting for real issues to expose project-specific knowledge gaps, \approach synthesizes project-specific issues by re-implementing test-covered core functionalities of the repository. By resolving these synthetic issues, \approach distills reusable project-specific knowledge into entity-grounded skills and associates them with relevant repository entities for future issue resolution.
Extensive experiments using both open-source and closed-source models show that \approach consistently improves issue resolution performance over strong baselines. These results demonstrate that proactively acquiring project-specific knowledge before solving real issues substantially improves downstream software issue resolution\footnote{Our code and data are available at https://github.com/cslsolow/SkillForge}.

\end{abstract}

\begin{IEEEkeywords}
Software engineering agents, software issue resolution, large language models
\end{IEEEkeywords}

\begin{figure*}[!t]
  \centering
  \includegraphics[width=\textwidth]{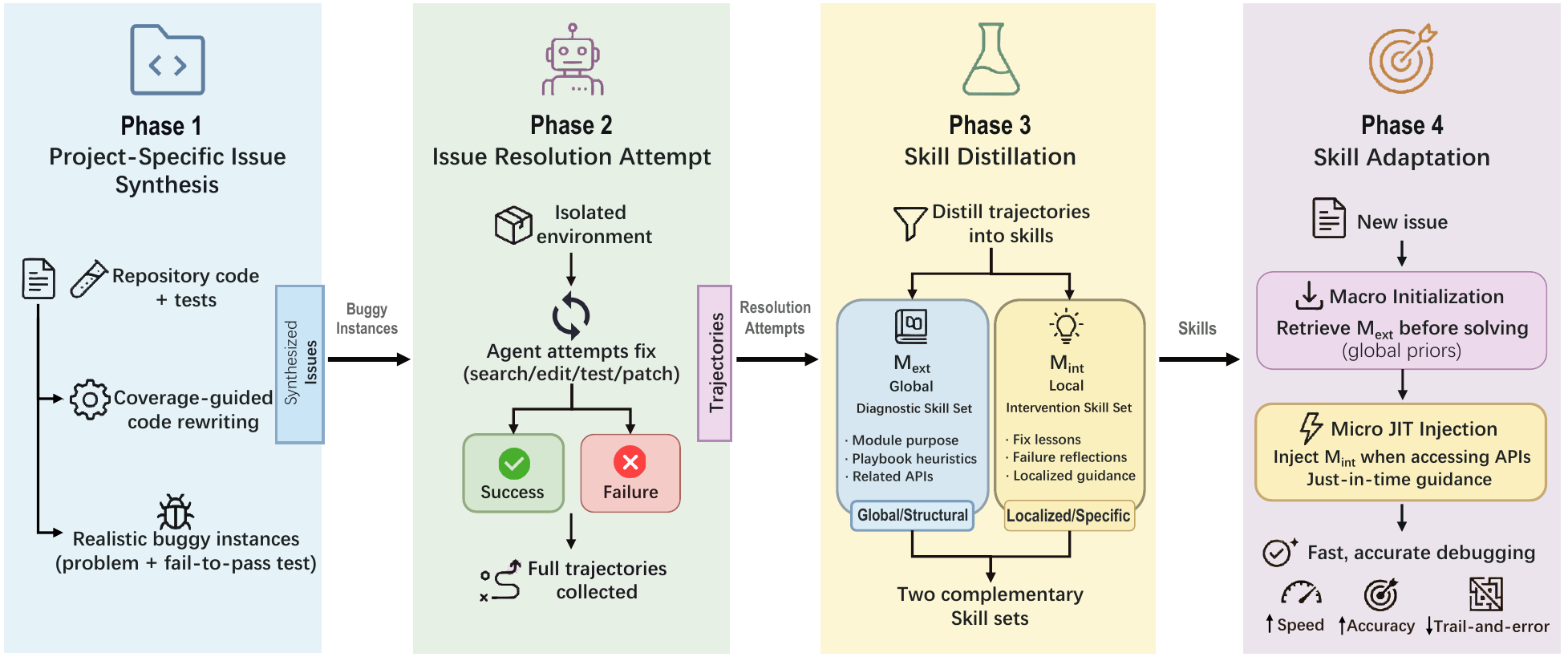}
  \caption{Overview of \approach.}
  \label{fig:framework}
\end{figure*}

\section{Introduction}

Large language model (LLM) based agents have demonstrated remarkable capabilities in software issue resolution~\cite{antoniades2024swesearch,chen2025swe,jimenez2023swe,ma2026llmagentscoderepositories,gao2025trae,kon2026swe, jiang2025putting,jiang2026koco,dong2025survey, gu2018deep,gu2016deep,jia2026compressing,ma2026samesignal,shi2026swebenchpromaxbenchmarkingagents,lin2026knowfixqadrivenrepository}. By autonomously navigating repositories, invoking tools, and iteratively proposing patches, modern software engineering (SWE) agents---such as SWE-agent~\cite{yang2024sweagenta} and OpenHands~\cite{wang2025openhandsopenplatformai}---can resolve complex, real-world bugs that previously required expert human intervention. Driven by increasingly powerful foundation models and better scaffolding designs, these agents have achieved impressive results on standard benchmarks such as SWE-bench~\cite{jimenez2023swe,oliva2025spice,wang2025solvedissues,ahmed2025testoverfitting,yu2026sweabs,yu2025utboost,garg2025savingswebench}, demonstrating strong generalization across diverse codebases and issue types.

Despite these advances, a critical bottleneck emerges when agents are deployed on a specific project: they typically must solve issues from scratch, without project-specific knowledge. Real-world projects often exhibit structural regularities: recurring failures may involve the same brittle modules, correct patches may need to preserve repository-specific API contracts, and related APIs are often coupled through implicit execution paths. Agents that lack such project-specific knowledge must repeatedly rediscover these conventions during issue resolution. This limitation creates a cold-start problem: before project-specific knowledge has been acquired, even a highly capable agent is reduced to a generic repository explorer and can repeatedly fall into the same project-specific pitfalls.

To address this knowledge gap, recent works have explored self-evolving SWE agents that acquire project-specific knowledge from prior issue-resolution signals~\cite{chen2025swe, lin2024llms, wang2026memgovern, mu2025experepair, hayashi2025self, li2025swe, xia2025livesweagent}. History-driven methods such as EvoCoder~\cite{lin2024llms} and SWE-Exp~\cite{chen2025swe} distill reusable knowledge from historical issues, commits, or agent trajectories. Online methods instead refine the agent during test-time exploration by generating additional trajectories on the current issue. While effective, both paradigms acquire project-specific knowledge reactively. History-driven methods are bounded by the richness and coverage of past issues: repository behaviors, API combinations, and long-tail components that have not been exercised by historical trajectories provide little learning signal. Online methods reduce this dependence on history, but they pay substantial trajectory, token, and time costs for each target issue~\cite{fan2025sweeffi}, and the acquired guidance becomes available only after the issue has already arrived.


In this paper, we propose \approach, a proactive self-distillation framework for addressing the cold-start problem in project-specific issue resolution. Rather than waiting for real issue-resolution history to accumulate, \approach synthesizes project-specific issues from the current repository. Given a repository snapshot, \approach starts from test-covered core functionalities, follows execution traces to identify the code regions that jointly implement the same functionality, and rewrites these coordinated segments under constrained context. The resulting synthetic issues are executable and behaviorally grounded by the repository tests. A SWE agent then resolves these synthetic issues, producing trajectories from which \approach distills project-specific knowledge.

\approach organizes the distilled knowledge as a dual-level skill repository. The global diagnostic skill set ($\mathcal{M}_{ext}$) captures reusable project-level knowledge for diagnosis and navigation, including entity roles, reasoning playbooks, and related APIs. The local intervention skill set ($\mathcal{M}_{int}$) records entity-specific modification guidance and pitfall-avoidance lessons distilled from successful and failed resolution trajectories. During downstream issue resolution, \approach first retrieves relevant global diagnostic skills as initial project-specific context, and then injects local intervention skills just in time whenever the agent accesses corresponding repository entities. This design treats skills as one concrete representation of project-specific knowledge, while keeping retrieval aligned with the agent's current code interaction context.

We implement \approach with Mini-SWE-Agent and evaluate it on SWE-bench Verified~\cite{openai2024swebenchverified} and SWE-bench Pro~\cite{deng2025swe} using both DeepSeek-V3.2 and GPT-5-mini. On SWE-bench Verified, \approach achieves 72.2\% and 60.6\% Pass@1, improving over Mini-SWE-Agent by +5.8 and +5.6 percentage points, respectively. On SWE-bench Pro, \approach further improves over Mini-SWE-Agent by +5.8 and +4.1 percentage points. Across both benchmarks, \approach outperforms all evaluated history-driven and online project-specific knowledge acquisition baselines available for each benchmark, showing that proactive project-specific knowledge acquisition can improve issue resolution without relying on rich historical issue trajectories or heavy per-issue online exploration.

The main contributions of this paper are summarized as follows:
\begin{itemize}
    \item We propose a novel self-distillation paradigm for addressing the cold-start problem in project-specific issue resolution. \approach proactively acquires project-specific knowledge from a repository's own tests and code, without relying on rich historical issue-resolution trajectories or costly per-issue online exploration.
    \item We design a dual-level skill repository that represents project-specific knowledge as global diagnostic skills and local intervention skills, enabling entity-grounded retrieval and just-in-time guidance during downstream issue resolution.
    \item Empirically, \approach improves over Mini-SWE-Agent by +5.8\%/+5.6\% on SWE-bench Verified and +5.8\%/+4.1\% on SWE-bench Pro for DeepSeek-V3.2 and GPT-5-mini, respectively, outperforming all evaluated history-driven and online project-specific knowledge acquisition baselines available for each benchmark.
\end{itemize}

\section{Methodology}

\approach is a self-distillation framework for addressing the cold-start problem in project-specific issue resolution. The key idea is to proactively derive project-specific knowledge directly from the repository itself, rather than waiting for real-world issue-resolution trajectories to accumulate. To achieve this, \approach synthesize project-specific issues, resolves them with a SWE agent, and distills the resulting project-specific knowledge into reusable skills that can be retrieved during subsequent real-world issue resolution.

Figure~\ref{fig:framework} presents an overview of \approach. Given a repository snapshot and its test suite, the framework consists of four stages: (1) generating project-specific synthetic issues by rewriting core functionalities; (2) resolving these issues with a SWE agent through iterative patch generation, tool execution, and test feedback; (3) distilling project-specific knowledge from the resulting resolution trajectories and organizing it as reusable skills in a dual-level skill repository, comprising a global diagnostic skill set that captures reusable project-level knowledge and a local intervention skill set that records entity-specific guidance; and (4) retrieving the relevant skills during downstream real-world issue resolution whenever the corresponding repository entities are encountered. Each stage is described in detail below.

\subsection{Project-Specific Issue Synthesis}

To acquire project-specific knowledge without relying on historical issue reports, \approach synthesizes project-specific issues by rewriting functionality-critical code segments and treating the resulting test failures as resolution targets. Rather than explicitly injecting handcrafted bugs, the framework asks an LLM to reimplement repository functionality under constrained context, encouraging the rewritten code to naturally exhibit implementation mistakes similar to those introduced during real software development. More importantly, because the model must reconstruct functionality without access to the original implementation, the rewritten code reflects its general coding knowledge rather than the repository's project-specific knowledge, thereby exposing the project-specific behaviors that the agent needs to learn. Throughout this process, supervision is derived solely from the repository's codebase and test suite, without relying on historical issue reports or human-written bug descriptions.

\textbf{Test-driven scope and trace.}
Given a code repository $\mathcal{R}$, \approach identifies test cases that exercise core repository functionalities, since these behaviors provide the most informative supervision for project-specific knowledge acquisition~\cite{chang2026test,chen2025oldtests}. For each passed test case, \approach executes it under coverage instrumentation to obtain an execution trace: the set of source files and line ranges that are exercised. This trace defines the candidate code regions for rewriting since modifications within executed regions are more likely to affect observable test outcomes. We then extract contiguous code segments from the traced regions---either complete functions or cohesive execution fragments, and attach surrounding context (e.g., preceding and following lines) for each segment. 

\textbf{Select critical segments.}
The number of traced segments can be large. We use an LLM to select a small set of \emph{critical} segments that are most likely to represent the functionality of this test case. The LLM is given the test's purpose and scenario (derived from the test code or description) along with summaries of the candidate segments (file, line range, and code content). It returns a ranked subset of top-$k$ segments with brief justifications. This step keeps the subsequent rewriting phase focused on high-impact regions. Unlike function-level rewriting approaches, \approach may select multiple critical segments exercised by the same test case, enabling synthetic issues that capture cross-component interactions within the repository.

\begin{figure}[t]
  \centering
  \includegraphics[width=\columnwidth]{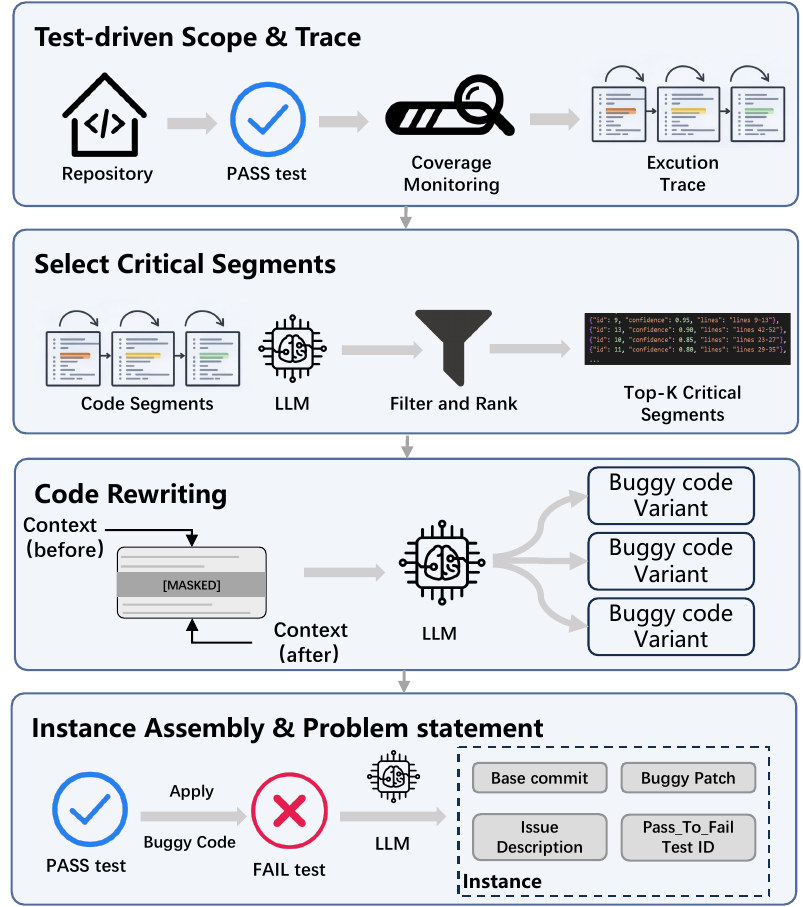}
  \caption{Synthesize project-specific issues. }
  \label{fig:phase1}
  \vspace{-8pt}
\end{figure}

\textbf{Code rewriting.}
For each selected segment, we ask an LLM to \emph{rewrite} the corresponding code by re-completing the local functionality. Crucially, the LLM is not provided with the original implementation; instead, it receives only limited contextual information: the lines before and after the segment, the segment's location and indentation, and a high-level description of the test's goal. The model is prompted to produce a plausible alternative implementation that preserves the intended API while potentially simplifying logic. Rather than explicitly injecting predefined faults, this ``strict-mask'' design exposes the gap between an agent's general coding knowledge and the repository's project-specific knowledge by inducing realistic implementation mistakes under constrained context.

\textbf{Instance assembly and problem statement.}
Once the rewritten segments induce test failures, we construct a buggy repository snapshot and derive both a buggy patch and a reference patch. Following prior synthetic SWE task construction works~\cite{yang2025swe,sonwane2025bugpilot,jain2025r2egym}, we execute the failing tests and use an LLM to convert the resulting failure evidence into a user-facing problem statement without exposing implementation details or repair hints. The resulting synthetic instance consists of a base commit, a buggy patch, a reference patch, a problem statement, and the associated test cases, matching the standard SWE-bench issue-resolution format. These synthetic instances serve as self-supervised tasks for acquiring project-specific knowledge before real issue resolution.

\textbf{Issue Resolution Attempt}
For each synthesized instance, \approach follows the SWE-bench-style issue resolution protocol: the agent is placed in an isolated repository environment, receives the problem statement and buggy codebase, and attempts to produce a patch that resolves the failing tests. We record the resulting action trajectory (e.g., file edits, shell commands, and test runs) for subsequent skill distillation.

\subsection{Skill Distillation}
\label{sec:distill}

The objective of this stage is to distill project-specific knowledge from synthetic issue-resolution trajectories and organize it as reusable skills. Formally, given a trajectory $\mathcal{T} = \{(a_1, o_1), \dots, (a_n, o_n)\}$ where $a_i$ and $o_i$ denote the agent's action and the environment's observation at step $i$, our extraction pipeline transforms $\mathcal{T}$ into two complementary skill sets within the dual-level skill repository: $\mathcal{M}_{ext}$ and $\mathcal{M}_{int}$.

\textbf{Trajectory Normalization and Entity Alignment.}
Before skill distillation, we normalize the trajectories and align them with the repository's entities. Given a target repository $\mathcal{R}$, we perform a structured analysis of the trajectory $\mathcal{T}$ to identify code-access events. By parsing shell commands (e.g., \texttt{grep}, \texttt{sed}, \texttt{cat}) in the trajectory, we extract a set of accessed files, represented by coordinates and line ranges. We then align these coordinates with the repository's structural index---an AST-derived mapping from source files to their class and function hierarchies---to resolve them into cohesive code scopes (e.g., frequently accessed files or core modules). Beyond trajectory-derived accesses, files surfaced in the agent's submission diff, the golden patch, and the test patch are additionally incorporated to ensure coverage of ground-truth edit sites. The resulting collection is denoted as a set of candidate entities $\mathcal{E}_{cand} \subseteq \mathcal{R}$. This step ensures that the extracted skills are strictly grounded in the actual codebase structure, preventing the LLM from hallucinating non-existent interfaces during subsequent extraction.

\begin{figure}[t]
  \centering
  \includegraphics[width=\columnwidth]{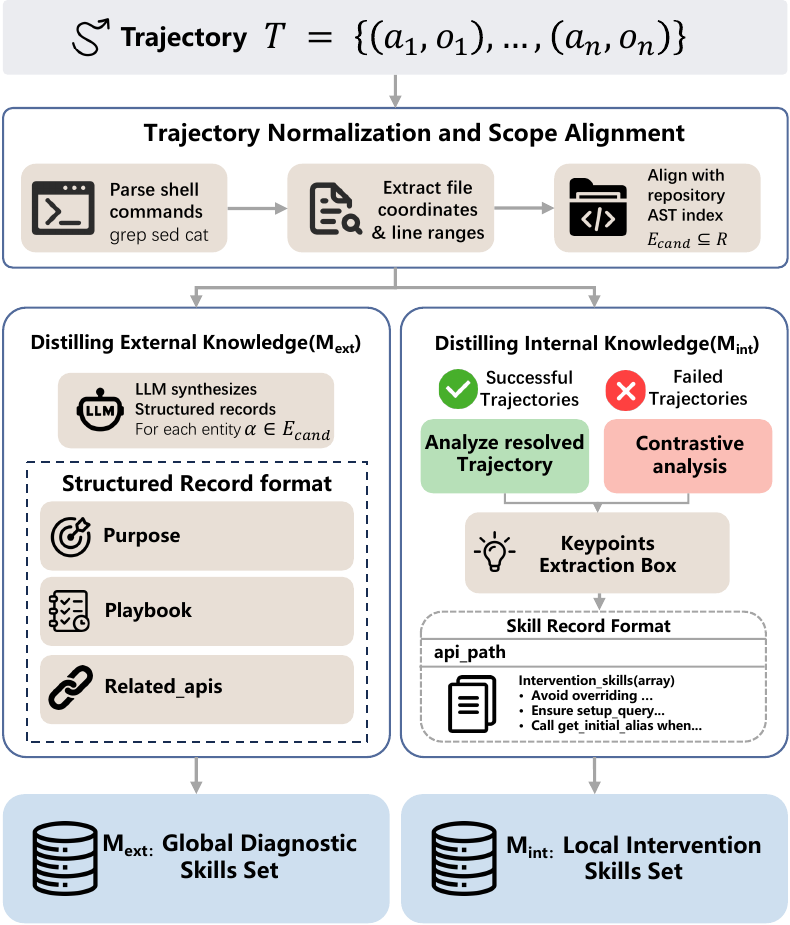}
  \caption{Skill distillation.} 
  \label{fig:phase3}
\end{figure}

\textbf{Distilling Global Diagnostic Skills ($\mathcal{M}_{ext}$).}
$\mathcal{M}_{ext}$ represents project-specific diagnostic knowledge as global diagnostic skills distilled from synthetic issue-resolution trajectories. Unlike repository summaries that describe what a module contains, $\mathcal{M}_{ext}$ captures how an agent should reason about a repository entity during issue resolution, including where to begin debugging, which APIs are jointly involved, and what repository-specific behaviors should be considered. These skills are organized as structured skill records and associated with the corresponding repository entities for future retrieval. For each candidate entity $\alpha \in \mathcal{E}_{cand}$, an LLM analyzes the corresponding resolution trajectories together with the repository context to distill three complementary aspects of project-specific knowledge: (i) the entity's functional role in issue resolution, (ii) reusable reasoning strategies repeatedly validated during issue resolution , and (iii) project-specific API interactions that emerge during problem solving. These are represented as the \texttt{purpose}, \texttt{playbook}, and \texttt{related\_apis} fields, respectively. For example,

\begin{lstlisting}
{
  (*@\textcolor{keycolor}{"api\_path"}@*): (*@\textcolor{valuecolor}{"django/db/models/sql/compiler.py"}@*),
  (*@\textcolor{keycolor}{"purpose"}@*): (*@\textcolor{valuecolor}{"Compiles Django ORM queries into SQL statements, ..."}@*),
  (*@\textcolor{keycolor}{"playbook"}@*): (*@\textcolor{valuecolor}{"If the issue involves SQL ..., first check if setup\_query properly ..."}@*),
  (*@\textcolor{keycolor}{"related\_apis"}@*): [
    {(*@\textcolor{keycolor}{"api\_path"}@*): (*@\textcolor{valuecolor}{"django/db/models/sql/query.py"}@*),
      (*@\textcolor{keycolor}{"reason"}@*): (*@\textcolor{valuecolor}{"Provides the Query class that manages alias ..."}@*)},
    ...]
}
\end{lstlisting}

The design of each key is specifically tailored to enhance the agent's autonomous capabilities:
(1) The \texttt{purpose} field identifies the role an entity plays during issue resolution, helping the agent determine whether it is a relevant debugging entry point rather than merely summarizing its implementation.
(2) The \texttt{playbook} field captures project-specific reasoning patterns distilled from trajectories. Rather than generic debugging advice, it records repository-specific knowledge that has been validated through interaction with the repository.
(3) The \texttt{related\_apis} field records APIs that are repeatedly co-involved during issue resolution. Unlike static dependency graphs, these relationships reflect repository-specific interaction patterns observed during problem solving, allowing the agent to navigate cross-module behaviors that are not evident from structural dependencies alone.

Unlike repository summarization, which captures static code semantics, our objective is to distill actionable project-specific knowledge that emerges only when an agent attempts to solve project-specific issues.

\textbf{Distilling Local Intervention Skills ($\mathcal{M}_{int}$).}
In contrast, $\mathcal{M}_{int}$ stores project-specific intervention knowledge that guides how a repository entity should be modified during issue resolution. Unlike $\mathcal{M}_{ext}$, which supports repository navigation and diagnosis, $\mathcal{M}_{int}$ captures concrete repair knowledge distilled from issue-resolution trajectories. This knowledge is represented as local intervention skills associated with individual repository entities.
We distill this intervention knowledge from both successful and failed resolution trajectories. Successful trajectories reveal repair strategies that consistently lead to correct implementations, while failed trajectories expose repository-specific pitfalls by contrasting incorrect patches with the corresponding reference patches. The resulting knowledge is transformed into actionable intervention skills and indexed by repository entities. Combining successful and failed trajectories provides complementary supervision: the former reinforces effective repair patterns, whereas the latter helps the agent avoid recurring mistakes when modifying the same repository entities in future issue-resolution tasks. For example,

\begin{lstlisting}
{
  (*@\textcolor{keycolor}{"api\_path"}@*): (*@\textcolor{valuecolor}{"django/db/models/sql/compiler.py"}@*),
  (*@\textcolor{keycolor}{"intervention\_skills"}@*): [
    (*@\textcolor{valuecolor}{"Avoid overriding ... because ..."}@*),
    ...]
}
\end{lstlisting}
Conceptually, $\mathcal{M}_{ext}$ captures diagnostic skills for understanding repository entities and planning repository navigation, whereas $\mathcal{M}_{int}$ captures intervention skills for modifying those entities based on project-specific knowledge distilled from synthetic issue resolution.

\subsection{Skill Adaptation}


To utilize the acquired project-specific knowledge during downstream issue resolution, \approach retrieves the distilled skills associated with relevant repository entities and injects them into the agent's reasoning process. This retrieval follows a two-stage, context-aware mechanism consisting of macro-level initialization and micro-level just-in-time (JIT) intervention.

\textbf{Macro-level Initialization with Global Diagnostic Skills.}
Before the agent begins resolving a new issue, we first establish a global semantic context using the diagnostic knowledge represented by the global skill set ($\mathcal{M}_{ext}$). Given the new issue description as a query, we employ a BM25 retriever to identify the top-$k$ most relevant skill records from $\mathcal{M}_{ext}$. The selected records---comprising the API paths, their purposes, and the associated playbooks---are prepended to the agent's initial prompt as project-specific priors. This equips the agent with an immediate, high-level understanding of the repository's architecture and relevant previously distilled heuristics.

\textbf{Micro-level JIT Injection of Local Intervention Skills.}
While the macro-level injection provides initial guidance, flooding the context with all local intervention skills ($\mathcal{M}_{int}$) at once would introduce significant noise. Instead, \approach dynamically injects $\mathcal{M}_{int}$ based on the agent's real-time actions. At each interaction step, \approach monitors the agent's executed shell commands to extract the specific file paths it is currently accessing. If an accessed file matches an entry in $\mathcal{M}_{int}$, the corresponding intervention cues and reflections are appended to the agent's context as an auxiliary observation for the subsequent reasoning step. This dynamic mechanism ensures that the agent receives targeted, pitfall-avoidance guidance exactly when it navigates to a relevant codebase region, ensuring the injected skills are strictly aligned with the agent's current code interaction context.

Unlike existing methods that retrieve semantically similar records from a centralized knowledge store, \approach grounds project-specific knowledge directly to repository entities. Consequently, retrieval is triggered by the agent's interaction with the corresponding code entities rather than solely by semantic similarity. This entity-grounded design ensures that diagnostic and intervention knowledge is delivered precisely when the agent reaches the relevant repository context, reducing retrieval ambiguity while maintaining tight alignment between the injected knowledge and the code under inspection.

\section{Experimental Setup}
\subsection{Research Questions}
We aim to evaluate \approach by answering four research questions (RQs):

\textbf{RQ1 (Effectiveness of \approach):} What is the impact of \approach on issue resolution performance compared to other self-evolving methods?

\textbf{RQ2 (Ablation Study):} How do the different components of \approach individually affect its effectiveness?

\textbf{RQ3 (Impact of Hyperparameters):} How do the key hyperparameters influence the overall performance of \approach?


\textbf{RQ4 (Efficacy across Diverse Repositories):} Does \approach consistently demonstrate effectiveness across various repositories?

\subsection{Datasets}
We evaluate \approach on \textbf{SWE-bench Verified}~\cite{openai2024swebenchverified}, a human-validated benchmark released by OpenAI, which has been widely adopted as the standard for evaluating software engineering agents. The benchmark consists of 500 tasks sourced from popular Python repositories on GitHub. 

We additionally evaluate on the full \textbf{SWE-bench Pro}~\cite{deng2025swe}, consisting of 731 instances across Python, JavaScript, TypeScript, and Go repositories. This benchmark contains more challenging long-horizon software engineering tasks that better reflect realistic multi-file issue resolution.

\subsection{Baseline Methods}
We compare \approach with representative approaches that acquire project-specific knowledge from different supervision sources.

\textbf{History-driven project-specific knowledge acquisition.}
These methods acquire project-specific knowledge from historical issue-resolution issues, and reuse the distilled knowledge to assist future issue resolution.
\begin{itemize}
    \item \textbf{SWE-Exp}~\cite{chen2025swe}: A framework that distills diagnostic patterns and repair strategies from prior agent trajectories, utilizing a dual-agent architecture to provide actionable guidance for new tasks.
    \item \textbf{EvoCoder}~\cite{lin2024llms}: A multi-agent continual learning framework that uses trajectory-based reflection to progressively refine problem-solving strategies based on previously resolved cases.
    \item \textbf{MemGovern}~\cite{wang2026memgovern}: MemGovern converts human debugging traces from GitHub into structured memory, enabling code agents to leverage prior problem-solving knowledge for improved issue resolution.
\end{itemize}

\textbf{Online project-specific knowledge acquisition.}
These approaches adapt agents during the resolution of the current issue by distilling project-specific knowledge from substantial online reasoning trajectories through test-time exploration:
\begin{itemize}
    \item \textbf{SAGE}~\cite{hayashi2025self}: A plan-learning method that distills reusable problem-solving guidance from ongoing agent attempts to iteratively improve performance on the current instance.
    \item \textbf{SWE-Debate}~\cite{li2025swe}: A multi-agent competitive framework that evolves stronger fix strategies through structured debate over competing reasoning trajectories.
    \item \textbf{Live-SWE-agent}~\cite{xia2025livesweagent}: A live self-evolving software agent that autonomously updates its own scaffold during runtime while solving software engineering tasks.
\end{itemize}

\textbf{Variants of \approach.}
To isolate the source of project-specific knowledge in \approach, we compare against two controlled variants while keeping the same downstream reused mechanism:
\begin{itemize}
    \item \textbf{\approach w/ SWE-Smith.} A controlled synthesis variant that replaces our functionality-level repository probing with SWE-Smith single-function rewriting.
    \item \textbf{\approach w/ LLM Summary.} A static repository-understanding variant that replaces trajectory-distilled project-specific knowledge with LLM-generated repository summaries.
\end{itemize}

\subsection{Metrics}

We evaluate \approach from both effectiveness and cost-efficiency perspectives.

\textbf{Pass@1} denotes the percentage of issues successfully resolved on the first attempt, following the evaluation protocol of~\cite{antoniades2024swesearch,yang2024sweagenta}. It directly measures the framework’s ability to generate correct patches without relying on repeated repair iterations.

\textbf{Avg Cost} represents the average end-to-end monetary cost per evaluated issue, measured in U.S. dollars~\cite{fan2025sweeffi}. The reported cost amortizes offline pre-computation over the evaluated instances and includes all stages of the framework pipeline, including synthetic issue generation, trajectory collection, skill distillation, and online issue resolution. MemGovern~\cite{wang2026memgovern} constructs a pre-constructed $\sim$150K-card experience base from public GitHub issues using GPT-5.1; since the associated preprocessing cost is unavailable, it is excluded from Avg Cost.

\subsection{Implementation Details}
We implement \approach with Mini-SWE-Agent~\cite{mini_swe_agent}, a bash-based agent scaffold~\cite{guo2026eet, hayashi2025self, wang2026swe}, and evaluate DeepSeek-V3.2~\cite{liu2025deepseek} and GPT-5-mini~\cite{singh2025openai} with default inference settings.


For SWE-bench Verified, synthesis is temporally isolated. For each target instance, we first roll back the repository before both the golden patch and golden test patch; on this rolled-back snapshot, an LLM localizes semantically relevant tests because extracting the full test suite from every snapshot is time- and compute-intensive. We synthesize issues only from this filtered subset, yielding 577 synthesized issues. \approach itself does not require historical issue-resolution data; in the temporal evaluation protocol, the agent may retrieve only skills distilled from synthesized issues in the same repository whose source commits predate the target instance's gold commit, ensuring that no current-instance or future information is used.


For the two variants, we extract relevant functions from the baseline agent's trajectory, then apply either SWE-Smith-style single-function rewriting or an LLM summary of how those functions are jointly used; both keep \approach's skill format and injection interface. Across all stages, temperature is 0, the action budget is 250 steps, BM25 retrieves top-$5$ skills, and Table~\ref{tab:main_results} reports Pass@1 averaged over three runs.

\begin{table}[]
\centering
\caption{Main results on SWE-bench Verified.}
\label{tab:main_results}
\resizebox{\linewidth}{!}{
\begin{tabular}{l l c c}
  \toprule
  \textbf{Method} & \textbf{Model} & \textbf{Pass@1} & \textbf{Avg Cost} \\
  \midrule
  \multirow{2}{*}{Mini-SWE-Agent}
  & DeepSeek-V3.2 & 66.4\% & \$0.049 \\
  & GPT-5-mini    & 55.0\% & \$0.031 \\
  \midrule
  \multicolumn{4}{l}{\textit{History-driven project-specific knowledge acquisition}} \\
  \midrule
  \multirow{2}{*}{SWE-Exp}
  & DeepSeek-V3.2 & 69.0\%\sigone \up{2.6\%} & \$0.090 \\
  & GPT-5-mini    & 56.6\%\sigone \up{1.6\%} & \$0.065 \\
  \multirow{2}{*}{EvoCoder}
  & DeepSeek-V3.2 & 67.0\% \up{0.6\%} & \$0.064 \\
  & GPT-5-mini    & 58.4\% \up{3.4\%} & \$0.052 \\
  \multirow{2}{*}{MemGovern}
  & DeepSeek-V3.2 & 69.2\%\sigone \up{2.8\%} & -- \\
  & GPT-5-mini    & 58.0\%\sigone \up{3.0\%} & -- \\
  \midrule
  \multicolumn{4}{l}{\textit{Online project-specific knowledge acquisition}} \\
  \midrule
  \multirow{2}{*}{SAGE}
  & DeepSeek-V3.2 & 67.2\% \up{0.8\%} & \$0.081 \\
  & GPT-5-mini    & 56.0\%\sigone \up{1.0\%} & \$0.052 \\
  \multirow{2}{*}{SWE-Debate}
  & DeepSeek-V3.2 & 68.2\% \up{1.8\%} & \$0.382 \\
  & GPT-5-mini    & 56.4\% \up{1.4\%} & \$0.167 \\
  \multirow{2}{*}{Live-SWE-agent}
  & DeepSeek-V3.2 & 67.0\% \up{0.6\%} & \$0.050 \\
  & GPT-5-mini    & 55.6\% \up{0.6\%} & \$0.042 \\
  \midrule
  \multicolumn{4}{l}{\textit{Variants of \approach}} \\
  \midrule
  \multirow{2}{*}{\approach w/ SWE-Smith}
  & DeepSeek-V3.2 & 68.0\% \up{1.6\%} & \$0.088 \\
  & GPT-5-mini    & 56.4\%\sigone \up{1.4\%} & \$0.071 \\
  \multirow{2}{*}{\approach w/ LLM Summary}
  & DeepSeek-V3.2 & 68.7\%\sigone \up{2.3\%} & \$0.069 \\
  & GPT-5-mini    & 54.4\% \down{0.6\%} & \$0.065 \\
  \multirow{2}{*}{\textbf{\approach}}
  & DeepSeek-V3.2 & \textbf{72.2\%}\sigone \upbf{5.8\%} & \$0.074 \\
  & GPT-5-mini    & \textbf{60.6\%}\sigone \upbf{5.6\%} & \$0.066 \\
  \midrule
  \multicolumn{4}{r}{\sigone: $p-value<0.05$.} \\
\end{tabular}
}
\vspace{-8pt}
\end{table}

\section{Results}

\subsection{RQ1: Effectiveness of \approach}
Table~\ref{tab:main_results} summarizes the main results on SWE-bench Verified. 
\approach achieves 72.2\% and 60.6\% Pass@1 with DeepSeek-V3.2 and GPT-5-mini, 
outperforming all baselines by absolute margins of \textbf{+5.8\%} and 
\textbf{+5.6\%} over Mini-SWE-Agent. Among history-driven project-specific knowledge acquisition methods, MemGovern is 
the strongest competitor (69.2\%/58.0\%), yet \approach still surpasses it by 
+3.0\%/+2.6\%. One possible reason is that history-driven project-specific knowledge acquisition methods are bounded by the coverage and quality of repository history: knowledge distilled from prior repository histories only reflects previously observed issue types, often remains coarse or instance-tied, and external debugging discussions improve coverage only by weakening repository specificity. Among online project-specific knowledge acquisition methods, \approach also surpasses SAGE (67.2\%/56.0\%) and SWE-Debate (68.2\%/56.4\%), with gains of +5.0\%/+4.6\% and +4.0\%/+4.2\%, respectively, while avoiding the high per-issue exploration cost of methods such as SWE-Debate.
For DeepSeek-V3.2, it also outperforms Live-SWE-agent (67.0\%, \$0.050) by +5.2\%.
Compared with Mini-SWE-Agent and history-driven project-specific knowledge acquisition methods, the main cost difference of \approach lies in the offline pre-computation stage, where it synthesizes project-specific issues and distills skills from their resolution trajectories. Therefore, the reported Avg Cost includes amortized offline pre-computation; once this repository-level skill repository is built, the online cost of resolving each real issue remains close to that of the underlying agent.

Table~\ref{tab:compact_results} further reports results on SWE-bench Pro. \approach achieves 34.1\% Pass@1 with DeepSeek-V3.2 and 51.7\% with GPT-5-mini, improving over Mini-SWE-Agent by +5.8\% and +4.1\%, respectively, with both gains statistically significant ($p$-value $<$ 0.05). It also outperforms the strongest available Pro baselines, exceeding Live-SWE-agent by +1.7\%/+2.6\% and SWE-Exp by +4.7\%/+3.2\% under the two backbones.

Among the controlled variants, \approach w/ SWE-Smith reaches 68.0\%/\$0.088 and 56.4\%/\$0.071 under DeepSeek-V3.2 and GPT-5-mini, respectively, while \approach w/ LLM Summary reaches 68.7\%/\$0.069 and 54.4\%/\$0.065. These results show that skill injection alone is insufficient: both simply summarizing trajectory-extracted functions and rewriting one function at a time underperform the full method. SWE-Smith rewrites one function at a time, whereas \approach starts from a repository functionality exposed by a core test, follows its execution trace, and rewrites multiple coordinated functions or code segments that are naturally composed to realize that functionality. This functionality-level synthesis captures inter-function coordination, API coupling, and project-specific repair pitfalls, while more naturally surfacing the current LLM's repository-misaligned coding tendencies.

\begin{findingbox}{1}
  \approach outperforms all evaluated history-driven and online project-specific knowledge acquisition baselines available for each benchmark, improving over Mini-SWE-Agent by \textbf{+5.8\%}/\textbf{+5.6\%} on SWE-bench Verified and \textbf{+5.8\%}/\textbf{+4.1\%} on SWE-bench Pro.
\end{findingbox}

\begin{table}[t]
\centering
\caption{Main results on SWE-bench Pro.}
\label{tab:compact_results}
\resizebox{\linewidth}{!}{
\begin{tabular}{l cc cc}
\toprule
\multicolumn{1}{c}{} & \multicolumn{2}{c}{\textbf{DeepSeek-V3.2}} & \multicolumn{2}{c}{\textbf{GPT-5-mini}} \\
\cmidrule(lr){2-3}\cmidrule(lr){4-5}
\textbf{Method} & \textbf{Pass@1} & \textbf{Avg Cost} & \textbf{Pass@1} & \textbf{Avg Cost} \\
\midrule
Mini-SWE-Agent & 28.3\% & \$0.047 & 47.6\% & \$0.063 \\
SWE-Exp & 29.4\% \up{1.1\%} & \$0.083 & 48.7\% \up{0.9\%} & \$0.089 \\
Live-SWE-agent & 32.4\% \up{4.1\%} & \$0.051 & 49.1\%\sigone \up{1.5\%} & \$0.072 \\
\textbf{\approach} & 34.1\%\sigone \up{5.8\%} & \$0.069 & 51.7\%\sigone \up{4.1\%} & \$0.087 \\
\bottomrule
\multicolumn{5}{r}{\sigone: $p-value<0.05$.}
\end{tabular}
}
\vspace{-8pt}
\end{table}

\subsection{RQ2: Ablation Study}
We investigate two questions: (1) whether both forms of project-specific knowledge are necessary for issue resolution, and (2) whether project-specific knowledge distilled for one LLM backbone transfers to another.

\paragraph{Component ablation.}
Table~\ref{tab:ablation} reports the effect of removing each type of project-specific knowledge. Removing $\mathcal{M}_{ext}$ leads to drops of 3.8\% and 3.0\%, demonstrating the importance of repository-level diagnostic knowledge for issue resolution. Removing $\mathcal{M}_{int}$ causes slightly larger drops of 4.4\% and 3.4\%, showing that project-specific intervention knowledge distilled from synthetic issue-resolution trajectories is equally critical. These results indicate that effective use of project-specific knowledge requires both repository-level diagnostic knowledge and entity-level intervention knowledge, which provide complementary guidance throughout the issue-resolution process.

\begin{table}[t]
  \centering
  \caption{Ablation study results.}
  \label{tab:ablation}
  \begin{tabular}{l c c}
  \toprule
  \textbf{Approach} & \textbf{DeepSeek-V3.2} & \textbf{GPT-5-mini} \\
  \midrule
  w/o Global Diagnostic Skills & 68.4\% \textcolor{red}{($\downarrow$3.8\%)} & 57.6\% \textcolor{red}{($\downarrow$3.0\%)} \\
  w/o Local Intervention Skills & 67.8\% \textcolor{red}{($\downarrow$4.4\%)} & 57.2\% \textcolor{red}{($\downarrow$3.4\%)} \\
  \midrule
  \textbf{\approach} & \textbf{72.2\%} & \textbf{60.6\%} \\
  \bottomrule
  \end{tabular}
\end{table}

\begin{table}[t]
\centering
\caption{Cross-LLM skill transfer on SWE-bench Verified. Resolver LLM denotes the backbone used for real issue resolution, while Knowledge-source LLM denotes the backbone used to synthesize instances and distill skills.}
\label{tab:skill_transfer}
\begin{tabular}{l l c}
\toprule
\textbf{Resolver LLM} & \textbf{Knowledge-source LLM} & \textbf{Pass@1} \\
\midrule
\multirow{2}{*}{GPT-5-mini}
& GPT-5-mini & \textbf{60.6\%} \\
& DeepSeek-V3.2 & 55.0\% \\
\multirow{2}{*}{DeepSeek-V3.2}
& GPT-5-mini & 65.2\% \\
& DeepSeek-V3.2 & \textbf{72.2\%} \\
\bottomrule
\end{tabular}
\end{table}

\paragraph{Transferability of project-specific knowledge across LLM backbones.}
We further investigate whether the project-specific knowledge distilled by \approach is transferable across LLM backbones. Specifically, we use one LLM to synthesize project-specific issues, resolve them, and distill the resulting project-specific knowledge, which is represented as reusable skills. These skills are then injected during issue resolution performed by either the same or a different LLM. Table~\ref{tab:skill_transfer} exhibits a clear diagonal pattern: each resolver LLM achieves the best performance when paired with project-specific knowledge distilled from itself. DeepSeek-V3.2 reaches 72.2\% when using its own distilled skills but drops to 65.2\% when using GPT-5-mini's. Similarly, GPT-5-mini achieves 60.6\% with its own distilled skills but decreases to 55.0\% when using those distilled by DeepSeek-V3.2.
These results indicate that the distilled project-specific knowledge is not universally transferable across LLMs. Instead, it captures how a particular LLM reasons about and edits a specific repository. Since different LLMs exhibit different coding priors, the repository-specific mismatches exposed during synthetic issue resolution are likewise model dependent, leading to different project-specific knowledge being distilled. Consequently, knowledge distilled from one LLM provides limited guidance for another and may even introduce irrelevant context. 

\vspace{0.5em}
\begin{findingbox}{2} 
Effective issue resolution requires both repository-level diagnostic knowledge and entity-level intervention knowledge. Moreover, the distilled project-specific knowledge is LLM-specific rather than universally transferable across backbones.
\end{findingbox}
\vspace{0.5em}

\subsection{RQ3: Impact of Hyperparameters}

We investigate the sensitivity of \approach to two key hyperparameters governing project-specific knowledge acquisition and utilization: (1) the number of retrieved global diagnostic skill records per issue ($k_r$), and (2) the number of code segments rewritten to synthesize project-specific issues ($k_s$).

\begin{figure}[t]
  \centering
  \includegraphics[width=\linewidth]{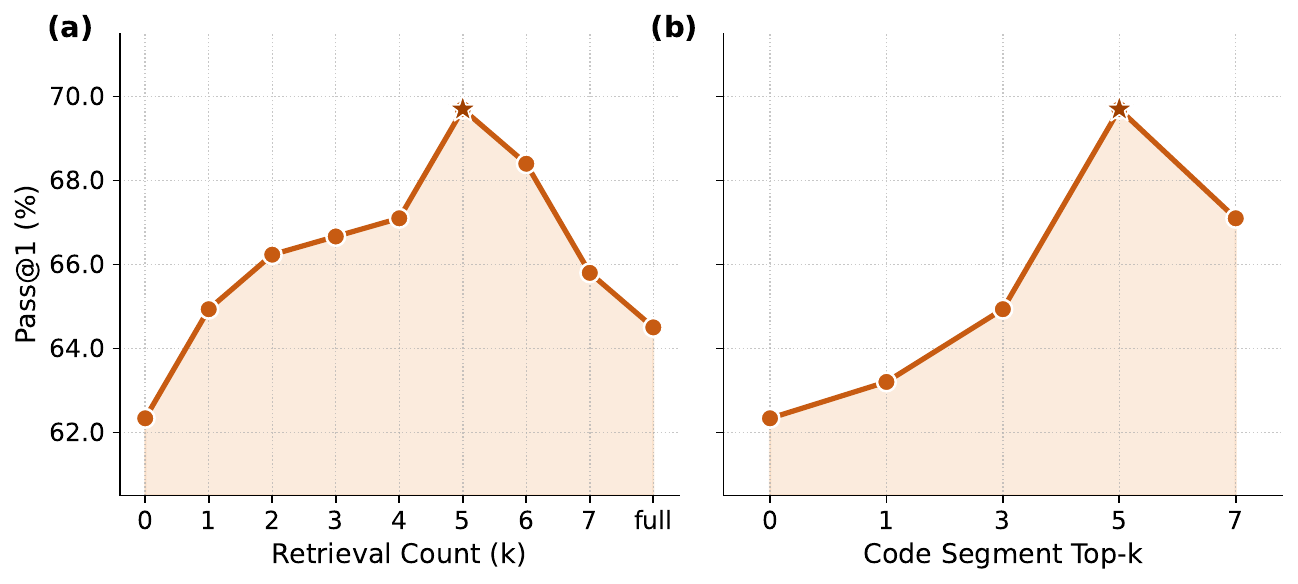}
  \caption{Hyperparameter study results on the Sphinx and Django repositories with GPT-5-mini. (a)~BM25 skill retrieval count. (b)~Code segment rewritten count during issue synthesis. Red stars mark the best-performing $k$ in each setting.}
  \label{fig:rq3}
  \vspace{-16pt}
\end{figure}

To evaluate $k_r$, we conducted a hyperparameter study on the Django and Sphinx repositories using GPT-5-mini, varying $k_r$ from 0 to 7, as well as a ``full'' condition where all relevant skill records were retrieved. Figure~\ref{fig:rq3}(a) reports Pass@1 as a function of $k_r$ across the two repositories. Without retrieving project-specific knowledge ($k_r=0$), the agent achieves 62.3\%. Performance improves steadily as $k_r$ increases, reaching its peak at $k_r=5$ (69.7\%). Beyond this point, the full retrieval condition drops slightly to 67.5\%, suggesting that lower-ranked skill records introduce redundant project-specific knowledge that competes for the limited context window. Both repositories show the same overall pattern, indicating that a moderate number of retrieved skills is preferable across different repository structures.

To evaluate $k_s$, we varied the number of code segments rewritten during issue synthesis on both repositories. With no rewriting ($k_s=0$), the system reduces to the baseline (62.3\%). Rewriting a single segment ($k_s=1$) increases performance to 63.2\%. Performance peaks at $k_s=5$, where each synthesized issue exposes sufficiently rich project-specific knowledge by involving multiple interacting repository entities. Increasing $k_s$ to 7 results in a slight decrease, as larger perturbations tend to generate overly complex issues that reduce the quality of the distilled project-specific knowledge. The same trend is consistently observed across repositories, suggesting that functionality-level synthesis benefits from a moderate rewriting scope rather than either single-segment perturbations or overly broad rewrites.

Notably, across all settings of $k_r$ and $k_s$, \approach consistently outperforms the baseline without project-specific knowledge, demonstrating that proactively acquired project-specific knowledge remains effective across a broad range of synthesis and retrieval configurations.





\begin{figure}[t]
  \centering
  \includegraphics[width=\linewidth]{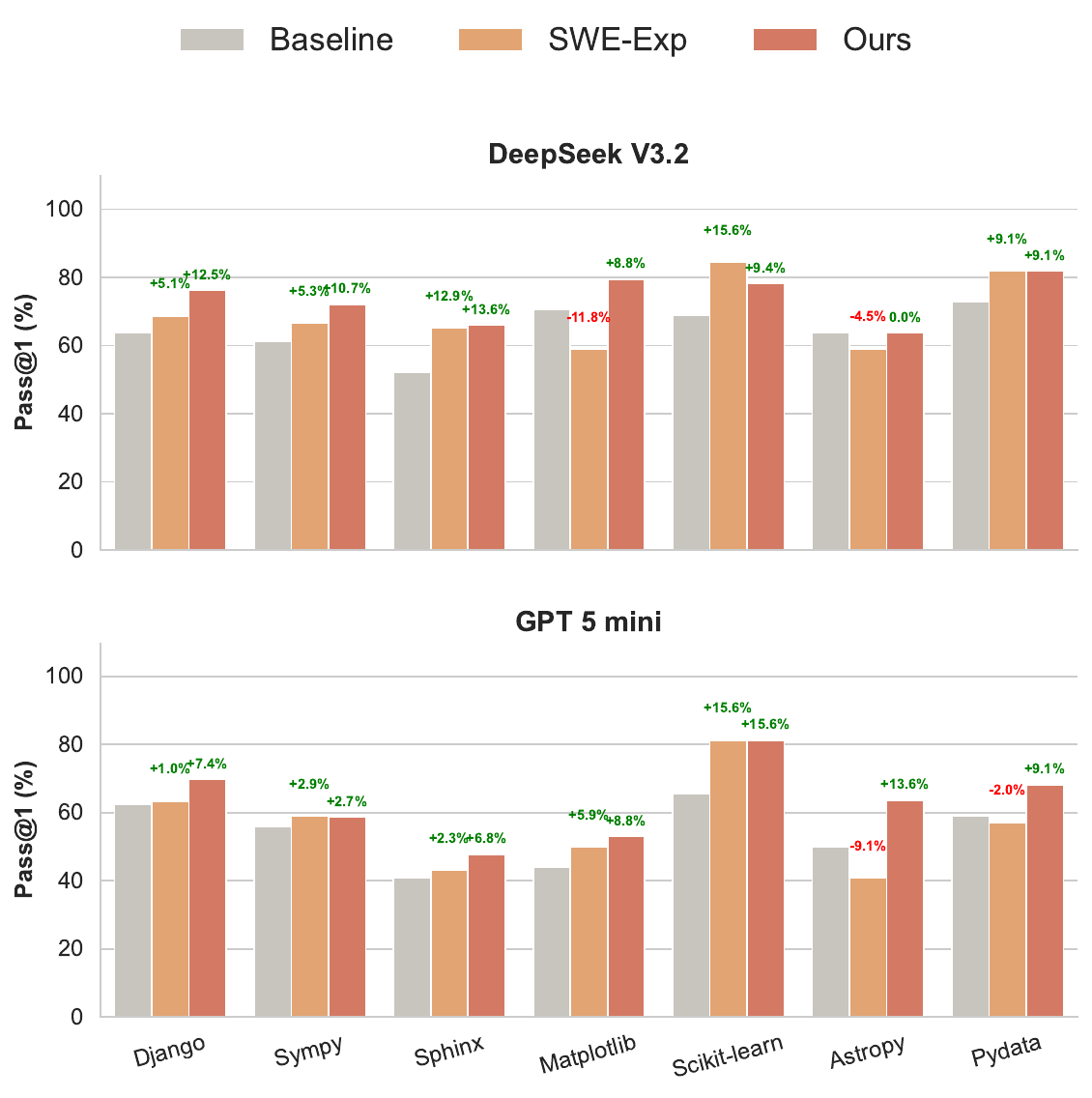}
  \caption{Pass@1 performance comparison between Baseline and our method across seven repositories.}
  \label{fig: diverse}
  \vspace{-16pt}
\end{figure}

\begin{findingbox}{3} 
Moderate synthesis and retrieval scales achieve the best performance, while excessive knowledge generation or retrieval introduces redundant information that slightly degrades performance.
\end{findingbox}
\vspace{0.5em}

\subsection{RQ4: Efficacy across Diverse Repositories}


To assess whether \approach generalizes across different repositories, we analyze performance at the repository level. Due to space constraints, we report results for the seven repositories with the largest numbers of evaluation instances. Figure~\ref{fig: diverse} compares three methods---Baseline, SWE-Exp, and \approach---across these seven repositories under both DeepSeek-V3.2 and GPT-5-mini.

\approach delivers improvements over the baseline on all seven repositories under both backbones without any regression. With DeepSeek-V3.2, the per-repo gains reach up to +13.6\% (Sphinx); under GPT-5-mini, gains reach up to +15.6\% (Scikit-learn).
In contrast, SWE-Exp shows inconsistent behavior and suffers regressions on three repositories. Under DeepSeek-V3.2, Matplotlib drops by $-$11.8\% and Astropy by $-$4.5\%; under GPT-5-mini, Astropy regresses by $-$9.1\% and Pydata by $-$2.0\%. 
Because \approach derives project-specific knowledge directly from each target repository through self-distillation, the resulting knowledge naturally reflects repository-specific APIs, implementation patterns, and architectural organization. Representing this knowledge as entity-grounded skills enables precise retrieval during downstream issue resolution, leading to consistently robust improvements across diverse repositories.


\vspace{0.5em}
\begin{findingbox}{4} 
\approach consistently improves Pass@1 across all evaluated repositories without regression, demonstrating robust gains from proactive project-specific knowledge acquisition.

\end{findingbox}

\subsection{Case Study}
\begin{figure*}[t]
  \centering
  \includegraphics[width=\textwidth]{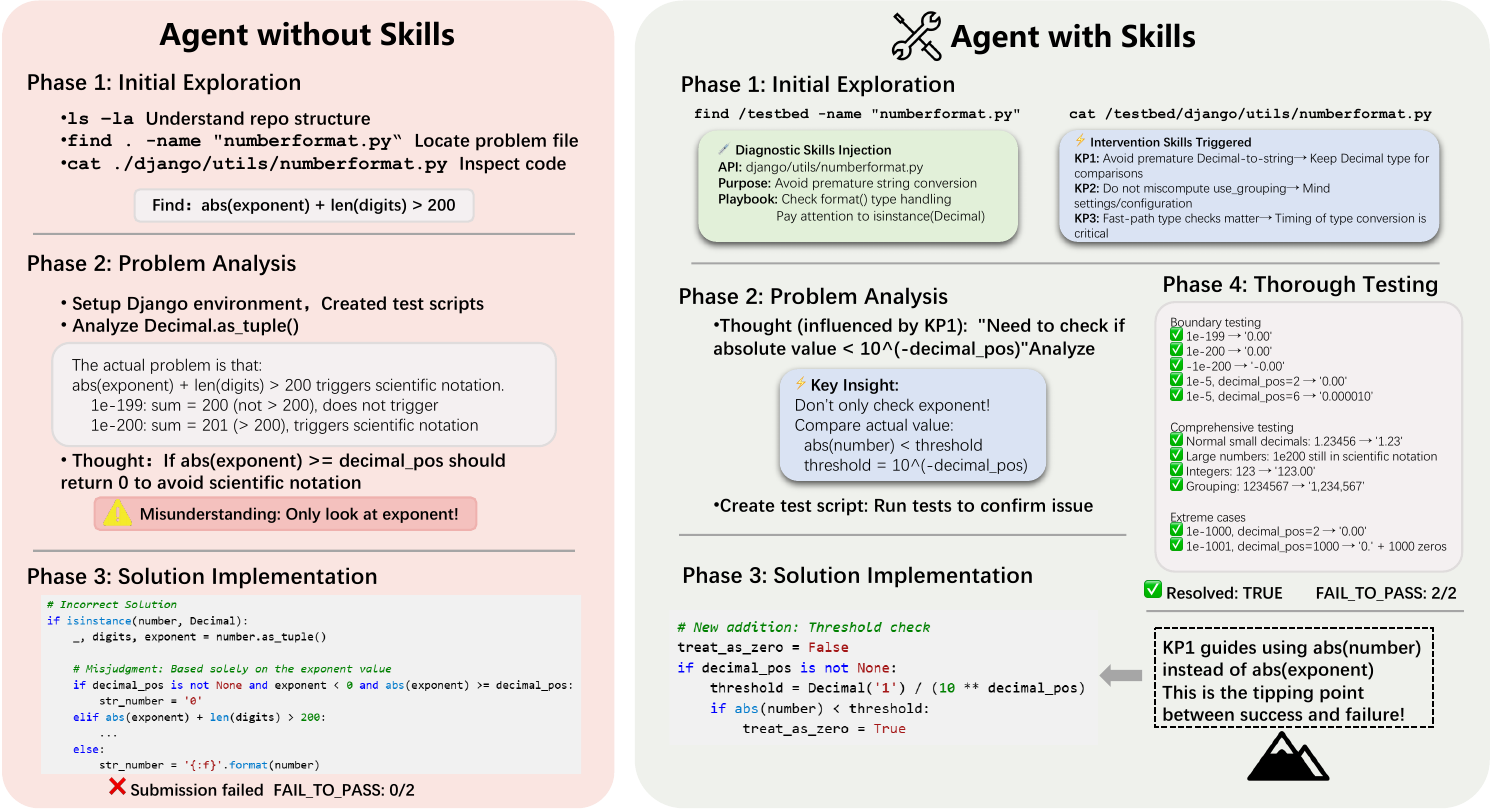}
  \caption{Case study for \textit{django-11206} with and without project-specifc knowledge (represented as skills).}
  \label{fig:case-study}
\end{figure*}

To provide a more concrete view of how \approach changes the behavior of a repair agent, we present a case study on Django issue \#11206. The bug concerns formatting extremely small \texttt{Decimal} values when a fixed number of decimal places is explicitly requested. In the buggy version, calling
\texttt{nformat(Decimal("1e-200"), ".", decimal\_pos=2)} returns \texttt{"1.00e-200"}, i.e., scientific notation, \allowbreak whereas users expect a fixed-point representation such as \texttt{"0.00"}. Correctly fixing this issue requires reasoning about when a value should be treated as numerically zero at a given precision, without breaking existing formatting behavior for other magnitudes or types.

Figure~\ref{fig:case-study} contrasts two repair trajectories on this issue using the same GPT-5-mini backbone: one without project-specific knowledge and one with \approach retrieving project-specific knowledge in the form of entity-grounded skills. The baseline agent locates the relevant formatting module and identifies the condition responsible for scientific notation. It then adopts a simple exponent-based heuristic to determine whether a value should be treated as zero. Although this modification fixes the observed failing example, it fails to capture the repository's intended numeric semantics and ultimately fails all FAIL\_TO\_PASS tests (0/2).

In contrast, the skill-enhanced agent follows a longer but more structured reasoning process. After accessing the same module, the retrieved project-specific knowledge highlights two repository-specific insights: (i) preserve the existing \texttt{Decimal} formatting pipeline instead of introducing alternative representations, and (ii) reason about numerical equivalence using the repository's precision semantics rather than a simple exponent heuristic. Guided by these insights, the agent derives a threshold-based solution that integrates naturally with the existing formatting logic. It further validates the implementation using representative boundary, precision-sensitive, and extreme-value test cases before committing the fix, ultimately passing all FAIL\_TO\_PASS tests (2/2).

This example illustrates how proactively acquired project-specific knowledge reshapes the agent's reasoning process during issue resolution. With skill injection, the agent (i) adopts a more faithful mathematical model of the bug (\emph{value-based thresholding} versus \emph{exponent-based heuristics}), (ii) preserves important design constraints of the existing implementation (e.g., maintaining \texttt{Decimal}-level operations and respecting the original formatting path), and (iii) naturally gravitates toward a test-driven workflow that exercises subtle edge cases. In contrast, the baseline trajectory quickly converges to a plausible but brittle local heuristic that fails under systematic evaluation, underscoring the value of skill-enhanced reasoning for non-trivial software bugs. 



\section{Discussion}

\subsection{Quality of Distilled Skills and LLM-Generated Problem Statements}
We manually inspected the quality of the distilled skills and synthesized problem statements. For the distilled skills, we verified that they are grounded in the corresponding repository entities, resolution trajectories, and golden patches, providing correct and actionable project-specific guidance without hallucinated APIs or misleading advice. For the synthesized problem statements, we verified that they faithfully describe the observable failures induced by functionality-level rewriting without leaking implementation details. 

We also compared the issue-type distribution of synthesized issues with that of real SWE-bench issues from the same repositories. The two distributions do not overlap, suggesting that the synthesized issues do not merely reproduce the same categories of real issues used for evaluation. Instead, they tend to expose finer-grained model biases in how the current LLM uses project-specific functions, APIs, and cross-function interactions. Thus, the benefit of \approach is less about teaching the agent to solve similar issue types and more about surfacing project-specific function-use pitfalls that can transfer across different real issue categories.

\subsection{Threats to Validity}
\paragraph{External validity.}
\approach relies on repository tests to synthesize issues and distill project-specific knowledge. Consequently, code regions that are rarely exercised by runnable tests contribute fewer learning signals, which may reduce the effectiveness of the learned knowledge in repositories with limited test coverage.

\paragraph{Internal validity.}
Following standard SWE-bench-style synthetic instance synthesis pipelines~\cite{jain2025r2egym, yang2025swe, sonwane2025bugpilot}, problem statements for synthesized issues are generated using an LLM. We manually verified that the synthesized failures correspond to executable bugs induced by functionality-level rewriting and that the generated descriptions correctly reflect observable behaviors without leaking implementation details. However, as in prior work, LLM-generated problem statements may still differ from developer-written issue reports.



\section{Related Work}

\subsection{Project-Specific Knowledge for SWE Agents}
Recent SWE agents~\cite{pan2026reporepair,team2026yet,raghavendra2026agentic,chen2026beyondswe,xu2025swe, zeng2025pruning,zeng2026dockerless,zeng2026glimprouter,hu2026line,gao2026swe,huang2027planning} have increasingly moved beyond one-shot issue solving toward \emph{self-evolving} behavior, where the agent improves by distilling reusable guidance from prior histories, trajectories, and execution feedback. These methods can be broadly grouped by the source of their evolution signal. One line of work evolves from repository history. EvoCoder~\cite{lin2024llms} and SWE-Exp~\cite{chen2025swe} accumulate past interaction trajectories and distill abstract problem-solving patterns to guide future tasks. ExpeRepair~\cite{mu2025experepair} organizes previously resolved bugs into structured memories for later patch generation, while MemGovern~\cite{wang2026memgovern} harvests community debugging traces from GitHub to improve the agent's decision-making process. Another line of work evolves from online trajectories produced on the current instance. SWE-Debate~\cite{li2025swe} iteratively strengthens solutions through multi-agent debate, SAGE~\cite{hayashi2025self} abstracts high-level guidance from trial-and-error grounding during test-time scaling, and Live-SWE-agent~\cite{xia2025livesweagent} evolves its own runtime scaffold while solving the current task. Unlike these reactive paradigms, \approach targets the cold-start problem in project-specific issue resolution. Instead of acquiring project-specific knowledge only after accumulating repository history or online exploration trajectories, it proactively derives such knowledge directly from the repository by synthesizing project-specific issues, and represents this knowledge as entity-grounded skills for future issue resolution.

\subsection{SWE-Style Instance Synthesis}
Recent work~\cite{yu2026survey,hu2025repo2run,wang2025swebenchplusplus} has explored synthesizing software issues or executable environments to improve software engineering agents. Existing methods primarily view synthesized instances as scalable training data. SWE-Smith~\cite{yang2025swe} constructs synthetic bug-fixing tasks by independently rewriting individual API-level functions. R2E-Gym~\cite{jain2025r2egym} generates executable SWE environments through test generation and commit back-translation to support large-scale agent training. BugPilot~\cite{sonwane2025bugpilot} synthesizes feature-development tasks that unintentionally introduce bugs, producing realistic development scenarios for subsequent model training.

In contrast, \approach uses synthetic issues for a fundamentally different purpose: proactive project-specific knowledge acquisition rather than training models. Instead of maximizing the number or diversity of synthesized tasks, \approach aims to expose project-specific knowledge that cannot be readily inferred from static repositories alone. Its issue construction also differs in granularity: rather than rewriting isolated API-level functions, randomly selecting several functions for joint rewriting, or generating new features, \approach starts from repository-tested functionalities and uses test execution traces to identify the code regions that jointly implement the same functionality. Solving the resulting issues reveals how the agent's general coding priors diverge from project-specific implementation patterns and API interactions.


\section{Conclusion}
We presented \approach, a self-distillation framework for addressing the cold-start problem in project-specific issue resolution. Rather than relying on repository history or costly per-issue test-time exploration, \approach proactively derives project-specific knowledge directly from the repository by synthesizing project-specific issues and distilling their resolution trajectories. This knowledge is represented as a dual-level, entity-grounded skill repository that supports future issue resolution. Experiments demonstrate that \approach consistently improves issue resolution performance, yielding absolute Pass@1 gains of +5.8\%/+5.6\% on SWE-bench Verified and +5.8\%/+4.1\% on SWE-bench Pro for DeepSeek-V3.2 and GPT-5-mini, respectively. Overall, our results suggest that proactively acquiring project-specific knowledge before real issue resolution is an effective and scalable alternative to reactive acquisition from repository history or online exploration.

\bibliographystyle{IEEEtran}
\bibliography{ref}

\end{document}